\documentclass{JHEP3} 

\JHEPspecialurl{http://jhep.sissa.it/JOURNAL/JHEP3.tar.gz}

\usepackage{epsfig,multicol,amsmath,amssymb,graphicx,bm}

\newcommand\fverb{\setbox\fverbbox=\hbox\bgroup\verb}
\newcommand\fverbdo{\egroup\medskip\noindent%
			\fbox{\unhbox\fverbbox}\ }
\newcommand\fverbit{\egroup\item[\fbox{\unhbox\fverbbox}]}
\newbox\fverbbox

\title{Skewness in CMB temperature fluctuations 
from curved cosmic (super-)strings}

\author{
Daisuke Yamauchi$^1$, Yuuiti Sendouda$^1$, Chul-Moon Yoo$^{1,2}$, Keitaro Takahashi$^3$, Atsushi Naruko$^1$, Misao Sasaki$^1$\\
	$^1$Yukawa Institute for Theoretical Physics, Kyoto University, Kyoto 606-8502, Japan\\
	$^2$Asia Pacific Center for Theoretical Physics, Pohang University of Science and Technology, Pohang 790-784, Korea\\
	$^3$Department of Physics and Astrophysics, Nagoya University, Nagoya 494-8602, Japan\\
	E-mail: \email{yamauchi@yukawa.kyoto-u.ac.jp, sendouda@yukawa.kyoto-u.ac.jp, 
yoo@yukawa.kyoto-u.ac.jp, keitaro@a.phys.nagoya-u.ac.jp, naruko@yukawa.kyoto-u.ac.jp, misao@yukawa.kyoto-u.ac.jp
}
	
}

\received{\today} 		
\accepted{\today}		

\preprint{YITP-10-9, APCTP Pre2010 - 002}	

\abstract{

We compute the one-point probability distribution function
of small-angle cosmic microwave background temperature 
fluctuations due to curved
cosmic (super-)strings with a simple model of string network 
by performing Monte Carlo simulations.
Taking into account of the correlation between the curvature and
the velocity of string segments, there appear 
non-Gaussian features, specifically non-Gaussian tails and a skewness,
in the one-point pdf. 
The obtained sample skewness for the conventional field-theoretic
cosmic strings is $g_1\approx -0.14$, which is consistent with the result 
reported by Fraisse et al.~\cite{Fraisse:2007nu}. 
We also discuss the dependence of the pdf on the intercommuting probability.
We find that the standard deviation of the Gaussian part 
increases and non-Gaussian features are suppressed as 
the intercommuting probability decreases.
For sufficiently small intercommuting probability, 
the skewness is given by $\lesssim \text{(a\ few)}\times 10^{-2}$.

}


\begin{document} 


\section{Introduction}

Cosmic strings are line-like topological 
defects formed in the early universe through 
a spontaneous symmetry breakdown.
It was claimed that the formation of cosmic strings
at the end of inflation is a generic feature of supersymmetric 
grand unified theories~\cite{Jeannerot:2003qv}. 
The string tension $\mu$ is directly related to 
the symmetry breaking energy scale.
Also recent studies of stringy cosmology have
revived interest in cosmic strings, because it was
pointed out that brane inflation models may produce 
another class of string objects, called cosmic 
superstrings~\cite{Sarangi:2002yt,Davis:2005dd,Copeland:2009ga,Majumdar:2005qc}.
Cosmic superstrings may have different
properties from those of 
conventional field-theoretic cosmic strings.
One of the observationally interesting differences 
is the intercommuting probability $P$.
It can be significantly smaller than unity for 
cosmic superstrings~\cite{Jackson:2004zg},
while normally $P=1$ for field-theoretic strings~\cite{Eto:2006db}
(but see \cite{Salmi:2007ah}).

The imprint of cosmic strings on the cosmic microwave 
background (CMB) has been widely studied.
Although cosmic strings were excluded as a dominant source of 
the observed large angular scale 
anisotropy ~\cite{Pogosian:2008am,Bevis:2007gh,Pogosian:2003mz}
(see also \cite{Perivolaropoulos:2005wa} for a review),
a signal due to cosmic strings could still be observed at small 
angular scales~\cite{Fraisse:2007nu,Pogosian:2008am} 
with future arcminutes experiments such as 
South Pole Telescope~\cite{Ruhl:2004kv}
or Atacama Cosmology Telescope~\cite{Kosowsky:2004sw}.
On small angular scales, the primary fluctuations are damped and 
only the integrated Sachs-Wolfe (ISW) effect is relevant.
The Gott-Kaiser-Stebbins (GKS) effect~\cite{Kaiser:1984iv,Gott:1984ef}
is the most characteristic signal of the ISW effect 
due to cosmic strings. 
The GKS effect is caused by a discontinuity of the
gravitational potential across a moving string segment
when photons passes by the moving string. 
If photons are scattered by a number of moving string segments, 
the observed temperature fluctuations appear as a superposition of
the discontinuities.

Since the network of strings is a highly nonlinear object,
its non-Gaussian features may help us 
distinguish cosmic string signals from other 
secondary effects and hence may enhance the observability.
A skewness, which is a measure of the asymmetry of a
probability distribution function, is one of the simplest features of 
non-Gaussianity~\cite{Komatsu:2001rj}.
The skewness of temperature fluctuations is defined by
\begin{eqnarray}
g_1=\frac{\overline{(\Delta -\bar{\Delta})^3}}{\sigma_{\Delta}^{3}}\,,
\label{eq:skew def}
\end{eqnarray}
where $\Delta \equiv (T-\bar{T})/T=\Delta T/T$, the bar `` $\bar{~}$ " denotes
the statistical average over a CMB map and $\sigma_{\Delta}$
the standard deviation.
Recently, Fraisse et al.~\cite{Fraisse:2007nu} 
found that the one-point probability distribution function (pdf) 
of the temperature fluctuations due to conventional cosmic strings 
has non-Gaussian tails and a negative skewness, $g_1\approx -0.23$.

Non-Gaussian features would also appear in the bispectrum~\cite{Gangui:2001fr}.
Some authors~\cite{Hindmarsh:2009qk,Hindmarsh:2009es,Regan:2009hv}
calculated the bispectrum and the trispectrum by using the formula 
derived by Hindmarsh~\cite{Hindmarsh:1993pu}. 
In particular, Hindmarsh et al.~\cite{Hindmarsh:2009qk}
pointed out that the numerically measured amplitude corresponds to
a $|f_{\rm NL}|\approx 10^3$ for the 
local-type of primordial non-Gaussianity.

In \cite{Takahashi:2008ui}, we computed analytically the one-point pdf 
of small-scale temperature fluctuations 
with a simple model of long straight segments and kinks.
We derived basic equations for the evolution of
string segments in cosmological backgrounds
by extending the velocity-dependent one-scale
model with the intercommuting probability 
$P\neq1$~\cite{Takahashi:2008ui,Martins:2000cs,Martins:1996jp,Avgoustidis:2005nv}. 
It was found that
the obtained one-point pdf consists of a Gaussian component due to
frequent scatterings by long straight segments and non-Gaussian
tails due to close encounters with kinks.
The dispersion of the Gaussian component is consistent with the result 
in \cite{Fraisse:2007nu} and the non-Gaussian tails also 
can be fitted to that in \cite{Fraisse:2007nu} by using 
two phenomenological parameters. 
The effect of the intercommuting probability $P$ was also 
studied there.
It was shown that the non-Gaussian tails diminish as $P$ decreases.
However the obtained one-point pdf is symmetric 
for positive and negative temperature fluctuations, hence did
not reproduce the non-zero skewness reported in \cite{Fraisse:2007nu}.

In this paper, we study in depth the non-Gaussian features,
especially skewness, of the CMB temperature fluctuations 
due to cosmic (super-)strings, and interpret numerical results 
obtained in \cite{Fraisse:2007nu}.
For this purpose, we adopt a simple model
of string network and perform Monte Carlo simulations.
We consider the correlation between the curvature
and the velocity of string segments as the origin of skewness 
and study the dependence of the skewness
on the intercommuting probability $P$. 

This paper is organized as follows.
In section \ref{sec:string dynamics},
we give a brief review of the derivation of 
basic equations for the evolution of string segments 
incorporating the intercommuting probability $P$~\cite{Takahashi:2008ui}.
In section \ref{sec:HSV formula and correlations}
we introduce general formulae for the temperature fluctuations
due to cosmic strings~\cite{Hindmarsh:1993pu,Stebbins:1994ng,Stebbins:1987va}.
We also discuss the correlation between 
the curvature and velocity of string segments
which gives rise to the skewness in the one point pdf.
Then, in section \ref{sec:numerical study}, we give the result
of our Monte Carlo simulations. We show
that the obtained one-point pdf has 
a negative skewness when the correlations are taken into account. 
The skewness is plotted as a function of $P$. 
Finally, we summarize our results in section \ref{sec:summary}.

\section{An analytic model of the string network}
\label{sec:string dynamics}

A string worldsheet can be 
described by $x^\mu=x^\mu(\sigma^a)$, 
where $x^\mu$ and $\sigma^a$ are the spacetime coordinates
and worldsheet coordinates, respectively. 
The induced metric of the worldsheet $\gamma_{ab}$ 
is given by 
\begin{eqnarray}
\gamma_{ab}
=g_{\mu\nu}\frac{dx^{\mu}}{d\sigma^{a}}
\frac{dx^{\nu}}{d\sigma^{b}}, 
\end{eqnarray}
where $g_{\mu\nu}$ is the spacetime metric.
We assume that the action for the string dynamics is 
well approximated by the Nambu-Goto action:
\begin{eqnarray}
S_{\rm NG}=\mu\int \sqrt{-\gamma}d^2\sigma\,,
\end{eqnarray}
where $\mu$ is 
the tension of the cosmic string. 

Let us consider string dynamics in
Friedmann-Lema\^itre-Robertson-Walker 
universes with the metric 
\begin{eqnarray}
ds_{\rm FLRW}^2=a^2(\eta )\left( -d\eta^{2} +d{\bm r}^2\right)\,. 
\end{eqnarray}
We choose the temporal gauge :
\begin{eqnarray}
\sigma^{0}=\eta\,,\ \sigma^{1}=\sigma\,,\ 
\dot{\bm r}\cdot{\bm r}^{\prime}=0\,,
\label{eq: temporal gauge}
\end{eqnarray}
where the bold letters denote the $3$-vectors on the comoving space
and the dot and the prime denote the derivative 
with respect to $\eta$ and $\sigma$, respectively.
The equations of motion on the cosmological 
backgrounds
are given by~\cite{Vilenkin-Shellard}
\begin{eqnarray}
&&\ddot{\bm r}+2{\cal H}
\left( 1-\dot{\bm r}^2\right)\dot{\bm r}
=\frac{1}{\epsilon}
\left(\frac{{\bm r}^{\prime}}{\epsilon}\right)^{\prime}
\,,
\label{eq:velocity EOM}\\
&&\dot{\epsilon}=-2{\cal H}\epsilon\dot{\bm r}^2\,,
\label{eq:energy EOM}
\end{eqnarray}
where 
${\cal H}\equiv \dot{a}/a$ 
and $\epsilon$ is the energy per unit coordinate 
length defined by
\begin{eqnarray}
\epsilon =\biggl[\frac{{{\bm r}^{\prime}}^2}
{1-\dot{\bm r}^2}\biggr]^{1/2}\,.
\end{eqnarray}

We define the total energy $E$ and the average
root-mean-square velocity of a string:
\begin{eqnarray}
&&E=\mu a(\eta )\int\epsilon d\sigma
\label{eq:defE}\,,\ 
v_{\rm rms}^2=\langle\dot{\bm r}^2\rangle
\equiv
\frac{\int\dot{\bm r}^2\epsilon d\sigma}{\int\epsilon d\sigma}
\,.
\end{eqnarray}
In the velocity-dependent one-scale model (VOS), 
the string network is characterized by only two physical quantities:
the correlation length $\xi$ and 
the rms velocity $v_{\rm rms}$.
The correlation length $\xi$ is defined by 
\begin{equation}
\xi \equiv\sqrt{\frac{\mu}{\rho}}\,, 
\label{eq:correaltion length}
\end{equation}
where $\rho$ is the total string energy density,
which related to $E$ as 
\begin{equation}
\rho\propto E/a^3\,. 
\end{equation}

In our treatment, we also take the energy loss due to 
loop formations into account.
A loop formation can occur through the intercommutation
of two segments or self-intercommutation of a single segment.
The characteristic time scale of the interval of 
loop formations with the intercommutation probability $P$ is 
$\sim\xi /(Pv_{\rm rms})$. 
Then the energy loss due to the loop formation 
can be described as 
\begin{equation}
\left(\frac{d\rho}{dt}\right)_{\rm to~loops}
=-\tilde c P v_{\rm rms}\frac{\rho}{\xi}\,, 
\label{eq:looploss}
\end{equation}
where we have introduced $\tilde c$ as a constant 
which represents the efficiency of loop formations. 

For brevity, we use the quantity $\gamma\equiv 1/H\xi$
rather than $\xi$ with $H=a{\cal H}$.
For a universe with the scale factor $a(t)\propto t^{\beta}$
with physical time $t\equiv \int a(\eta )d\eta$, the
equations of motion for $\gamma$ and $v_{\rm rms}$
can be derived from Eqs.~\eqref{eq:defE} - \eqref{eq:looploss} as
\cite{Takahashi:2008ui,Martins:2000cs,Martins:1996jp,Avgoustidis:2005nv}
\begin{eqnarray}
&&\frac{t}{\gamma}\frac{d\gamma}{dt}
=1-\beta -\frac{1}{2}\beta\tilde{c}Pv_{\rm rms}\gamma
-\beta v_{\rm rms}^2
\,,\label{eq:gamma eq}\\
&&\frac{dv_{\rm rms}}{dt}
=(1-v_{\rm rms}^2)
\biggl[ \frac{k}{R} -2Hv_{\rm rms}\biggr]\,, 
\label{eq:v_rms eq}
\end{eqnarray}
where we have replaced $\langle(\dot{\bm r}^2)^2\rangle$ 
by $\langle\dot{\bm r}^2\rangle^2$.
The momentum parameter $k$ and the curvature radius $R$ are
defined below.
For the curvature term, we set
\begin{eqnarray}
\frac{\partial^{2}{\bm r}}{\partial\hat{s}^2}
=\frac{a(\eta )}{R}\hat{\bm u}\,,
\end{eqnarray}
where we have introduced the proper coordinate length measure 
along a string by 
$d\hat{s}\equiv |{\bm r}^{\prime}|d\sigma 
=\sqrt{1-\dot{\bm r}^2}\epsilon d\sigma$
and $\hat{\bm u}$ is the unit vector 
parallel to the curvature radius vector 
on the comoving space. The momentum parameter 
$k$ is defined by
\begin{eqnarray}
&&k\equiv
\frac{\big\langle 
(1-\dot{\bm r}^2)(\dot{\bm r}\cdot\hat{\bm u})
\big\rangle}
{v_{\rm rms}(1-v_{\rm rms}^2)}
=\frac{R}{a(\eta )v_{\rm rms}(1-v_{\rm rms}^2)}
\bigg\langle
\dot{\bm r}\cdot\frac{1}{\epsilon}
\left(\frac{{\bm r}^{\prime}}{\epsilon}\right)^{\prime}
\bigg\rangle
\,.
\label{eq:k def}
\end{eqnarray} 
In this paper, we use an approximated form of $k$~\cite{Martins:2000cs}:
\begin{eqnarray}
k(v_{\rm rms})=\frac{2\sqrt{2}}{\pi}
\frac{1-8v_{\rm rms}^6}{1+8v_{\rm rms}^6}\,.
\end{eqnarray}

It is known that a string network approaches the 
scaling regime where the characteristic scale grows 
with the horizon size~\cite{Kibble:1984hp,Ringeval:2005kr}. 
For our Monte Carlo simulations, we assume that the scaling is 
already realized by the time photons leave the last scattering surface.
This means that $\gamma$ and $v_{\rm rms}$ are 
constant in time. Hence we set the left-hand sides of
Eqs.~(\ref{eq:gamma eq}) and (\ref{eq:v_rms eq}) equal to zero.
We solve them numerically with the assumptions that
the universe is in the matter-dominated era, $\beta =2/3$, 
and the curvature radius is equal to the correlation length, $R=\xi$.

\section{CMB Temperature Fluctuations}
\label{sec:HSV formula and correlations}

%

\subsection{The Hindmarsh-Stebbins-Veeraraghavan formula}
\label{sec:HSV formula}

A formula for the temperature fluctuation
due to a Nambu-Goto string was derived by Stebbins 
and Veeraraghavan~\cite{Stebbins:1994ng,Stebbins:1987va}
and by Hindmarsh~\cite{Hindmarsh:1993pu}. 
The Hindmarsh-Stebbins-Veeraraghavan (HSV) 
formula is given by 
\begin{eqnarray}
&&\Delta ({\bm n}\,,{\bm r}_{\rm obs}\,,\eta_{\rm obs})
\equiv\frac{\Delta T}{T}
=-2G\mu\int_{\Sigma} d\sigma
\frac{\left( 1+{\bm n}\cdot\dot{\bm r}\right)
\bigl[\left( X{\bm n}+{\bm X}\right)\cdot{\bm u}\bigr]}
{\left( X+{\bm n}\cdot{\bm X}\right)
\left( X-\dot{{\bm r}}\cdot{\bm X}\right)}
\biggl|_{\eta =\eta_{\rm lc}(\sigma )}
\,,\label{eq:general temp fluc}
\end{eqnarray}
where ${\bm X} (\sigma )\equiv {\bm r}_{\rm obs}-{\bm r}(\sigma ,\eta_{\rm lc}(\sigma ))$
is the comoving position of the observer 
relative to that of the string, $X\equiv |{\bm X}|$, 
${\bm u}$ is defined by
\begin{eqnarray}
&&{\bm u}\equiv\dot{{\bm r}}
-\left(\frac{{\bm n}\cdot{\bm r}^{\prime}}
{1+{\bm n}\cdot\dot{{\bm r}}}\right)
{\bm r}^{\prime}\,,
\label{eq:u}
\end{eqnarray}
and $\eta_{\rm lc}(\sigma )$ is the conformal time along the intersection
of the observer's past light-cone and the string worldsheet,
\begin{eqnarray}
\eta_{\rm obs}-\eta_{\rm lc}(\sigma )=X(\sigma )\,.
\label{eq:light-cone time}
\end{eqnarray}
It should be stressed that the temperature fluctuation depends 
on the string distribution on the observer's past light-cone.

The integration range $\Sigma$ must be 
appropriately determined.
In our calculation, we consider only a segment of a long cosmic string
at each scattering. Namely we take $\Sigma$ as 
\begin{eqnarray}
\Sigma\ :\ X^{\perp}(\sigma )< \frac{\xi}{a}\,,
\label{eq:Sigma def}
\end{eqnarray}
where
\begin{equation}
\ {\bm X} ^{\perp}(\sigma )\equiv
{\bm X}(\sigma ) -\left({\bm X}(\sigma ) \cdot{\bm n}\right)
{\bm n}\,.
 \end{equation}
This procedure is consistent 
with the treatment in our previous paper~\cite{Takahashi:2008ui}.

It is useful to adopt the small angle approximation 
to understand various effects contained in the HSV formula.
In the small-angle approximation,
we assume that the angle between 
the direction of the source $-{\bm X}$ and 
the line-of-sight ${\bm n}$ is sufficiently small. 
This situation is realized when the length of
 a string segment, which is approximately
equal to the Hubble radius at the epoch of scattering, 
is much smaller than the distance between
 the string and the observer.
Thus we have
\begin{eqnarray}
&&\frac{X^{\perp}(\sigma )}{X(\sigma )}\ll 1\,,\ \ 
{\bm X}(\sigma ) 
\approx -X{\bm n}+{\bm X} ^{\perp}
+\frac{\left( X^{\perp}\right)^{2}}{2X}{\bm n}
+\cdots
\,.\label{eq:small angle expansion}
\end{eqnarray}
This approximation is valid as far as a scattering due 
to a segment at large redshift $z\gg 1$ is considered.

Let us consider the leading order 
in the small angle approximation,
\begin{eqnarray}
&&\Delta\approx
-4G\mu\int_{\Sigma} d\sigma
\frac{{\bm X} ^{\perp}\cdot{\bm u}}{(X^{\perp})^2}
\biggl|_{\eta =\eta_{\rm lc}(\sigma )}
\,.\label{eq:temp fluc in small angle}
\end{eqnarray}
For convenience,
we decompose this into two parts as~\cite{Stebbins:1994ng}
\begin{eqnarray}
&&\Delta\approx\Delta^{\parallel}+\Delta^{\perp}
\,,\\
&&\Delta^{\parallel}\equiv 4G\mu\int_{\Sigma} d\sigma
\alpha_{\parallel}
\frac{{\bm X} ^{\perp}\cdot 
\frac{d{\bm X} ^{\perp}}{d\sigma}}
{(X^{\perp})^2}\Biggl|_{\eta_{\rm lc}(\sigma )}
\,,\label{eq:Delta^parallel def}\\
&&\Delta^{\perp}\equiv 4G\mu
\int_{\Sigma} d\sigma
\frac{|\dot{\bm r}|}{\sqrt{1-\dot{\bm r}^2}}
\alpha_{\perp}
\frac{\left({\bm n}\times {\bm X} ^{\perp}\right) 
\cdot \frac{d{\bm X} ^{\perp}}{d\sigma}}
{(X^{\perp})^2}\Biggl|_{\eta_{\rm lc}(\sigma )}\,,
\end{eqnarray}
where the coefficients $\alpha_\parallel$ and $\alpha_\perp$
are defined by
\begin{eqnarray}
&&\alpha_{\parallel}
=\frac{{\bm n}\cdot{\bm r}^{\prime}}
{|{\bm r}^{\prime}|^2}
\,,\quad 
\alpha_{\perp}
={\bm n}\cdot
\left(\frac{{\bm r}^{\prime}}{|{\bm r}^{\prime}|}
\times \frac{\dot{\bm r}}{|\dot{\bm r}|}\right)
\,.
\end{eqnarray}
Note that $d/d\sigma$ in the above formulae
is the derivative along 
the intersection of the past light-cone with the worldsheet,
\begin{eqnarray}
\frac{d}{d\sigma}
\equiv\frac{\partial}{\partial\sigma}
+\frac{d\eta_{\rm lc}(\sigma )}{d\sigma}
\frac{\partial}{\partial\eta}\,.
\label{dsigmadef}
\end{eqnarray}

We may rewrite Eq.~(\ref{eq:Delta^parallel def}) by integration
by part as
\begin{eqnarray}
&&\Delta^{\parallel}
=-4G\mu\int_{\Sigma} d\sigma
\frac{d\alpha_{\parallel}}{d\sigma}
\ln\left(\frac{X^{\perp}}{L_{\rm cut}}\right)
\biggl|_{\eta_{\rm lc}(\sigma )}
\,,
\label{eq:Delta^parallel}
\end{eqnarray}
where $L_{\rm cut}$ is the cutoff scale of the scattering,
$L_{\rm cut}=\xi/a$. It is easy to see from this expression
that for an exactly straight and uniformly moving segment, 
that is for $\dot{\bm r}={\rm const.}$ and ${\bm r}^{\prime}={\rm const.}$, 
$\Delta^{\parallel}$ vanishes.

If the segment has a kink at $\sigma =\sigma_{k}$, we have
\begin{eqnarray}
\alpha_{\parallel}=\alpha_{\parallel}^{(+)}\Theta (\sigma -\sigma_{k})
+\alpha_{\parallel}^{(-)}\Theta (\sigma_{k}-\sigma )\,,
\end{eqnarray}
where $\alpha^{(\pm)}_{\parallel}$ are the constant parameters characterizing the kink and
$\Theta (\sigma )$ denotes the Heaviside step function.
Then, $\Delta^{\parallel}$ reproduces the kink temperature fluctuation,
\begin{equation}
\Delta^{\parallel}\approx\Delta_{\rm kink}
=-4G\mu\alpha_{\rm kink}\ln\frac{X^{\perp}(\sigma_{k})}{L_{\rm kink}}\,,
\label{eq:deltakink}
\end{equation}
where we have defined
$\alpha_{\rm kink}=\alpha_{\parallel}^{(+)}-\alpha_{\parallel}^{(-)}$
and $L_{\rm kink}$ is a mean distance between kinks
\cite{Takahashi:2008ui,Stebbins:1987va,Stebbins:1994ng}.
If the curvature of the segment is 
taken into account, $d\alpha_{\parallel}/d\sigma\neq0$
and $\Delta^{\parallel}$ appears.
Therefore, $\Delta^{\parallel}$ can be regarded as 
contributions of the deficits on the segment 
(kinks and cusps) and the curvature of the segment.

For an exactly straight and uniformly moving segment,
the position of the segments, ${\bm X}^{\perp}(\sigma )$, 
can be written as
\begin{eqnarray}
{\bm X}^{\perp}(\sigma )={\bm \delta}
+\sigma\biggl|\frac{d{\bm X}^{\perp}}{d\sigma}\biggl|\,{\bm e}\,,
\end{eqnarray}
where we have introduced the impact parameter 
${\bm\delta}\equiv{\bm X}^{\perp}(\sigma =0)$,
$\delta\equiv |{\bm \delta}|$, 
and the unit tangent vector ${\bm e}\propto d{\bm X}^{\perp}/d\sigma$. 
Then $\Delta^{\perp}$ becomes
\begin{eqnarray}
&&\Delta^{\perp}
\approx -4G\mu\frac{|\dot{\bm r}|}{\sqrt{1-\dot{\bm r}^2}}
\alpha_{\perp}
\Biggl[
\arctan\Biggl(\frac{\bigl|\frac{d{\bm X}^{\perp}}{d\sigma}\bigl|\sigma 
+\delta\cos\varphi}
{\delta\sin\varphi}\Biggr)
\Biggr]^{\sigma_{+}}_{\sigma_{-}}
\nonumber\\
&&
\qquad
\approx -8G\mu\frac{|\dot{\bm r}|}{\sqrt{1-\dot{\bm r}^2}}
\alpha_{\perp}
\arctan\Biggl(\frac{\xi}{a\,\delta\sin\varphi}\Biggr)\,,
\label{eq:generalized GKS}
\end{eqnarray}
where ${\bm\delta}\cdot{\bm e}\equiv\delta\cos\varphi$.
It is clear to see that the dominant contributions are 
given from the string portions with 
$|\sigma |\lesssim\delta /\bigl|\frac{d{\bm X}^{\perp}}{d\sigma}\bigl|$.
This is just the Gott-Kaiser-Stebbins (GKS) 
effect~\cite{Kaiser:1984iv,Gott:1984ef}.
Hence $\Delta^{\perp}$ may be regarded as the generalization of the GKS effect.
If the curvature of the segment is taken into account, 
correction terms due to higher derivatives will appear in $\Delta^{\perp}$.

Although most scatterings occur at $z\gg 1$,
some photons may be scattered by a segment at $z\sim {\cal O}(1)$. 
Therefore it is desirable to check the validity of the small 
angle approximation. For this purpose, in our Monte Carlo simulations, 
we have taken into account the next order corrections 
and compared the result with the one obtained in the leading order 
approximation. We found no difference between these two, supporting
the validity of the small angle approximation.

\subsection{The Gott-Kaiser-Stebbins effect}
\label{sec:GKS effect and kinks}

Next we consider the GKS effect~\cite{Kaiser:1984iv,Gott:1984ef}.
Using Eq.~(\ref{eq:generalized GKS}), the rms temperature fluctuation 
due to the GKS effect of a string is estimated as
\begin{eqnarray}
\Delta_{\rm GKS}
=8\frac{v_{\rm rms}}{\sqrt{1-v_{\rm rms}^2}}\alpha_{\rm seg}G\mu
\arctan\left(\frac{\xi}{a\delta}\right)\,.
\label{eq:general GKS effect}
\end{eqnarray}
where $\alpha_{\rm seg}=\left\langle\alpha_\perp^2\right\rangle^{1/2}$.
In the case of a close encounter, i.e. $\delta\ll\xi/a$,
this reduces to
\begin{eqnarray}
\Delta_{\rm GKS}
\approx 4\pi\frac{v_{\rm rms}}{\sqrt{1-v_{\rm rms}^2}}\alpha_{\rm seg}G\mu
\,,
\label{eq.:GKS effect}
\end{eqnarray}
Thus, this produces the discontinuity of the temperature fluctuations
across the moving string segment
when photons passes by the moving string. 


\FIGURE{\epsfig{file=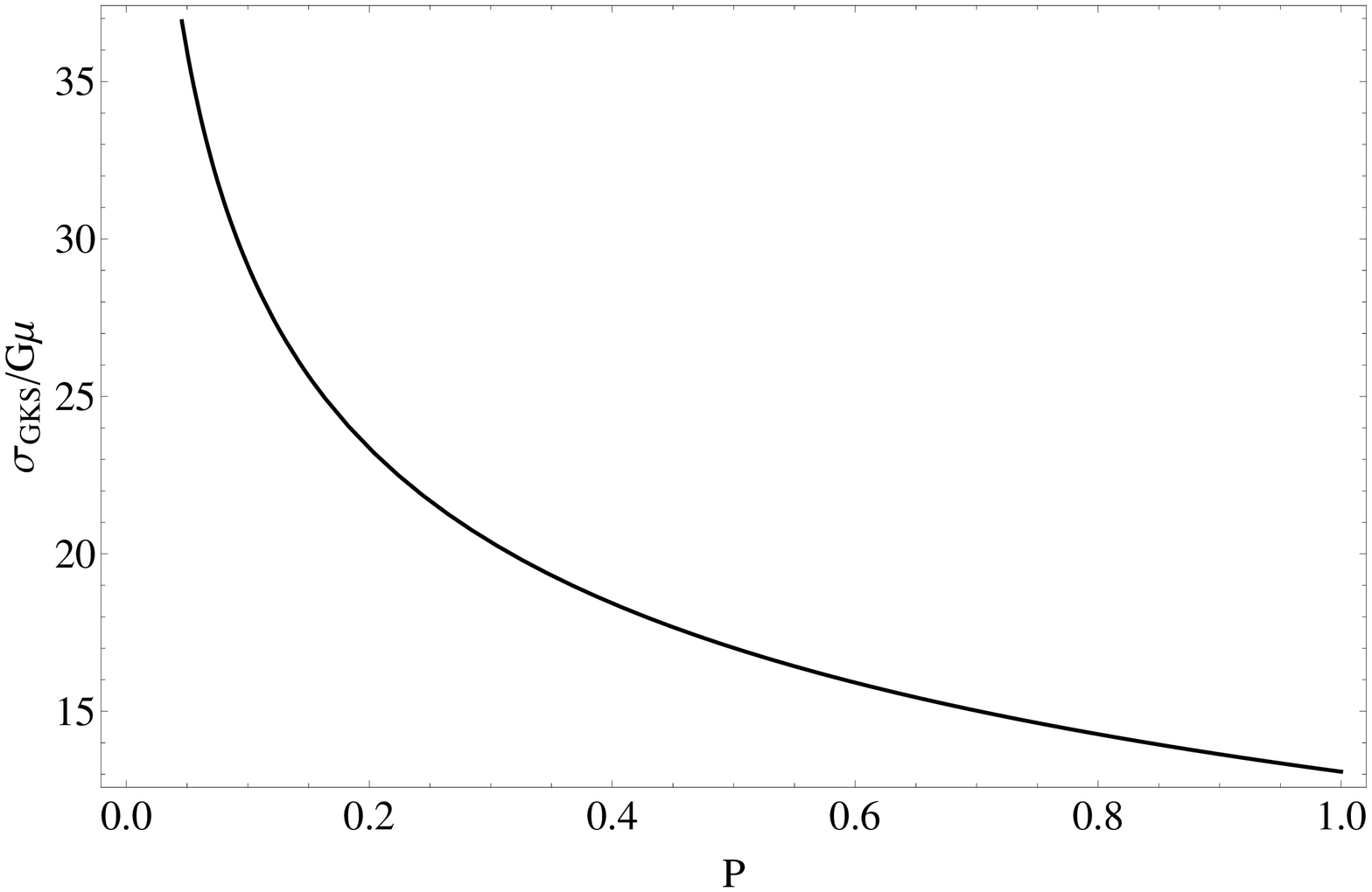,width=10cm} 
        \caption{The standard deviation of the Gaussian components of the pdf 
due to GKS effects, $\sigma_{\rm GKS}$
as a function of intercommuting probability $P$.}%
	\label{fig:sigma_seg}}

The temperature fluctuation due to a segment
approaches zero when the impact parameter is larger than
the segment length $\xi$. This implies that
a segment has an effective cross section $\sim\xi^{2}$. 
In terms of $\xi$, the number of segments 
in a horizon volume is expressed as $N_{\rm seg}=1/\xi^{3}H^3=\gamma^{3}$.
Therefore the optical depth for a CMB photon is
\begin{eqnarray}
\tau_{\rm opt}&=&\int_{0}^{z_{\rm LSS}}N_{\rm seg}H^3\xi^{2}
\frac{dz}{H(1+z)}
=\gamma\log (1+z_{\rm LSS})
\nonumber\\
&\approx& 16\left(\frac{\tilde{c}P}{0.23}\right)^{-1/2}
\,,\label{eq:optical depth}
\end{eqnarray}
where $\gamma\approx(3\tilde c P/\sqrt{2\pi^2})^{-1/2}$,
and we have put $z_{\rm LSS}\approx 1100$ and 
used $\tilde{c}\approx 0.23$ as the standard value~\cite{Martins:2003vd}. 
Since $\tau_{\rm opt}\gg1$,
a photon ray is scattered by segments many times
on its way from the last scattering surface 
to the observer.
Therefore, the temperature fluctuations behave 
like a random walk and the one-point pdf becomes well approximated
by a Gaussian distribution as a result of the central limit theorem.
Thus the GKS effect gives the Gaussian component of the one-point pdf,
\begin{eqnarray}
{\frac{dP}{d\Delta}}_{\rm GKS}
\approx\frac{1}{\sqrt{2\pi}
\sigma_{\rm GKS}}e^{-\Delta^{2}/2\sigma_{\rm GKS}^{2}}\,,
\end{eqnarray}
where the standard deviation is estimated as
\begin{eqnarray}
\sigma_{\rm GKS} 
=\Delta_{\rm GKS}\frac{\sqrt{\tau_{\rm opt}}}{2}
\approx 13G\mu
\left(\frac{\alpha_{\rm seg}}{1/\sqrt{2}}\right)
\left(\frac{\tilde{c}P}{0.23}\right)^{-1/4}\,,
\label{eq:sigma_GKS}
\end{eqnarray}
which is shown in Fig.~\ref{fig:sigma_seg}
as a function of intercommuting probability $P$.
In \cite{Takahashi:2008ui}, we found 
that the above estimate of the standard deviation 
agrees well with the result of numerical simulations (for $P=1$)
by Fraisse et al.~\cite{Fraisse:2007nu}.

\subsection{Correlations as sources of the skewness}
\label{sec:correlations}

The one-point pdf of the temperature fluctuations 
due to the GKS effect is symmetric under $\Delta \to -\Delta$ 
since the form of the GKS effect is obviously symmetric 
as is seen from Eq.~(\ref{eq:general GKS effect}).
However, a skewness may appear if we take account of
the curvature of segments and its correlations with the velocity,
as discussed in \cite{Takahashi:2008ui,Hindmarsh:2009qk}.
If a segment has a curvature, the probability of producing
positive or negative temperature deviation becomes asymmetric.
The amplitude of the asymmetry depends on the angle
between the curvature vector and the velocity vector 
of the segment. However, if the curvature and the 
velocity are not correlated, the statistical averaging
results in a vanishing skewness.
Therefore, a correlation between the velocity vector and 
the curvature vector is necessary to produce a skewness.
In this paper, we investigate two kinds of correlation. 
One is due to the breaking of the time reversal symmetry 
caused by the cosmic expansion, 
which has been already pointed out in \cite{Hindmarsh:2009qk}. 
The other is due to the effective curvature of a string 
on its intersection with the observer's past light-cone. 
We call this ``light-cone effect".

\subsubsection{Correlation due to cosmic expansion}

From Eq.~\eqref{eq:v_rms eq}, assuming
the scaling regime, we have 
\begin{eqnarray}
\frac{k}{R}=\frac{1}{av_{\rm rms}(1-v_{\rm rms}^2)}
\bigg\langle
\dot{\bm r}\cdot\frac{1}{\epsilon}
\left(\frac{{\bm r}^{\prime}}{\epsilon}\right)^{\prime}
\bigg\rangle
\approx 2Hv_{\rm rms}
\,.\label{eq:curvature-velocity correlation}
\end{eqnarray}
It is easy to see that a nontrivial correlation 
between the velocity vector $\dot{\bm r}$ and the curvature
vector ${\bm r}''$ exists if there is Hubble expansion. 
Each configuration of string segments must be
consistent with this correlation. 
We use this expression when we perform our Monte Carlo simulation.
A method for determining string segment configurations in our 
simulations is explained in section~\ref{sec:numerical study}.

\subsubsection{Light-cone effect}

In the HSV formula, Eq.~(\ref{eq:general temp fluc}),
the integration is along the light-cone. 
In the small angle approximation, the coordinates of a string
on the past light-cone emanating from the observer, ${\bm X}(\sigma)$,
can be expanded as
\begin{eqnarray}
&&{\bm X}(\sigma )
={\bm X}_0+\left(\frac{d{\bm X}}{d\sigma}\right)_{0}\sigma
+\frac{1}{2}\left(\frac{d^2{\bm X}}{d\sigma^2}\right)_{0}\sigma^{2}
+{\cal O}(\sigma^{3})
\nonumber\\
&&\ \ \ \ \ \ \ \ 
={\bm X}_0
-\Biggl[{\bm r}^{\prime}_0
+\left(\frac{d\eta_{\rm lc}}{d\sigma}\right)_{0}\dot{\bm r}_0\Biggr]
\sigma
\nonumber\\
&&\ \ \ \ \ \ \ \ \ \ 
-\frac{1}{2}\Biggl[{\bm r}^{\prime\prime}_0
+\left(\frac{d\eta_{\rm lc}}{d\sigma}\right)^{2}_0\ddot{\bm r}_0
+2\left(\frac{d\eta_{\rm lc}}{d\sigma}\right)_{0}
\dot{\bm r}^{\prime}_0
+\left(\frac{d^2\eta_{\rm lc}}{d\sigma^{2}}\right)_{0}\dot{\bm r}_0
\Biggr]\sigma^{2}
+{\cal O}(\sigma^{3})\,,
\label{eq:position expansion}
\end{eqnarray}
where $d\eta_{\rm lc}/d\sigma$ and $d^2\eta_{\rm lc}/d\sigma^{2}$ 
are estimated from Eqs.~(\ref{eq:light-cone time}) and 
(\ref{eq:small angle expansion}) as
\begin{eqnarray}
&&\left(\frac{d\eta_{\rm lc}}{d\sigma}\right)_{0}
=-\frac{{\bm n}\cdot{\bm r}^{\prime}_0}
{1+{\bm n}\cdot\dot{\bm r}_0}
\,,\label{eq:light cone time 1st}
\\
&&\left(\frac{d^2\eta_{\rm lc}}{d\sigma^{2}}\right)_{0}
=-\frac{1}{1+{\bm n}\cdot\dot{\bm r}_0}
\Biggl[ 
\frac{1-\dot{\bm r}^2_0}{X_0}
\Biggl\{
\frac{{{\bm r}^{\prime}_0}^2}{1-\dot{\bm r}^2_0}
-\left(\frac{d\eta_{\rm lc}}{d\sigma}\right)^{2}_0
\Biggr\}
\nonumber\\
&&\ \ \ \ \ \ \ \ \ \ \ \ \ \ \ \ \ 
+{\bm n}\cdot\Biggl\{
{\bm r}^{\prime\prime}_0
+\left(\frac{d\eta_{\rm lc}}{d\sigma}\right)^{2}_0
\ddot{\bm r}_0
+2\left(\frac{d\eta_{\rm lc}}{d\sigma}\right)_{0}
\dot{\bm r}^{\prime}_0
\Biggr\}
\Biggr]\,.
\label{eq:light cone time 2nd}
\end{eqnarray}

It is remarkable
that even if a string is exactly straight
and moves uniformly, 
${\bm r}^{\prime\prime}_0=\ddot{\bm r}_0=\dot{\bm r}^{\prime}_0=0$, 
it always has an ``effective curvature''.
Namely,
\begin{eqnarray}
\left(\frac{d^2{\bm r}}{d\sigma^2}\right)_0
=\left(\frac{d^2\eta_{\rm lc}}{d\sigma^2}\right)_0\dot{\bm r}_0
\neq{\bm 0}\,,
\end{eqnarray}
where note that $d/d\sigma$ is a 
derivative along the light-cone; see Eq.~(\ref{dsigmadef}).
This obviously means the existence of a correlation 
between the effective curvature and the velocity of the segment.
However, this effect is subdominant in the small angle approximation,
because of the suppression factor $\sigma/X_0$ 
in Eq.~\eqref{eq:position expansion}, where $X_0$ is of the
order of the angular diameter distance from the string
to the observer. 
Thus as far as scatterings at $z>1$ are concerned,
the light-cone effect is smaller than the effect due to
cosmic expansion.

\section{Monte Carlo simulations for the one-point pdf}
\label{sec:numerical study}

\subsection{Setup}
\label{sec:setup}

We compute the one-point pdf of the small-angle 
CMB temperature fluctuations due to curved string segments 
with a simple model of the string network (VOS)
by performing Monte Carlo simulations.

\subsubsection{Scattering probability}

Curved segments are assumed to be located
randomly between the LSS and the present time 
consistently with the VOS model.
We divide the redshift range $0<z<z_{\rm LSS}$ 
into a number of bins with the width 
\begin{equation}
\Delta\log (1+z)=\frac{\log (1+z_{\rm LSS})}{\tau_{\rm opt}}\Delta p 
=\frac{\Delta p}{\gamma}\,, 
\label{eq:scatteringprob}
\end{equation}
where $\Delta p$ is the scattering probability assigned to each bin. 
The scattering events due to cosmic strings can be randomly 
arranged on the observer's line-of-sight by using 
Eq.~\eqref{eq:scatteringprob}. 
We set $\Delta p=10^{-2}$ in our simulations. 

\subsubsection{Assumptions and the expansion of the HSV formula}
\label{subsubsec:hsvformula}

To take into account the curvature of a string segment,
we expand ${\bm X}^{\perp}(\sigma )$ to second order in $\sigma$
as given by (\ref{eq:position expansion}).
We assume the higher order terms are negligible. Hence we set
\begin{eqnarray}
\frac{{\bm X}^{\perp}(\sigma )}{X_0}=\frac{\mbox{\boldmath $\delta $}}{X_0}
+\left(\frac{d{\bm X}^{\perp}}{d\sigma}\right)_{0}\frac{\sigma}{X_0}
+\frac{1}{2}\left(\frac{d^2{\bm X}^{\perp}}{d\sigma^2}\right)_{0}
\frac{\sigma^{2}}{X_0}\,.
\label{eq:parabola shape}
\end{eqnarray}
We also assume the validity of the small angle approximation
and assume $\delta/X_0=O(\varepsilon)$ and 
$\bigl|\frac{d{\bm X}^{\perp}}{d\sigma}\bigl|_0\sigma /X_0=O(\varepsilon)$,
where $\varepsilon$ is a small expansion parameter.
Furthermore, we assume that the curvature ${\bm r}''_0\sigma$ , 
the rotation $\dot{\bm r}'_0\sigma$ and 
the acceleration $\ddot{\bm r}_0\sigma$ of 
a string segment can be treated as perturbations from a uniformly moving 
straight string segment,
\begin{eqnarray}
|{\bm r}''_0|\sigma\sim |\ddot{\bm r}_0|\sigma\sim |\dot{\bm r}'_0|\sigma
={\cal O}(\varepsilon)\ll 1\,.
\end{eqnarray}
The above order estimate comes from the fact that
${\bm r}''\sigma\sim \sigma/\xi\sim \sigma/X_0$. Note that
these order estimates imply that the first two, leading order terms
in (\ref{eq:parabola shape}) are already of $O(\varepsilon)$
and hence the last curvature term, which is $O(\varepsilon^2)$,
is only $O(\varepsilon)$ relative
to the leading order terms.

Now we can expand the reduced HSV formula 
Eq.~(\ref{eq:temp fluc in small angle}).
As discussed in the above, the effect of the curvature
appears $O(\varepsilon)$ relative to the leading order terms.
Therefore, we expand the numerator ${\bm X}^{\perp}\cdot{\bm u}$
to $O(\sigma^2)$ and the denominator $(X^{\perp})^2$ to $O(\sigma^3)$.
Hence it can be written as
\begin{eqnarray}
&&\Delta\approx -4G\mu\int_{\Sigma} d\sigma
\frac{\sum_{i=0}^{2}\alpha_{i}\sigma^{i}}
{\sum_{j=0}^{3}\beta_{j}\sigma^{j}}
\,,\label{eq:small angle expansion of HSV}
\end{eqnarray}
where $\alpha_{i}$ ($i=0,1,2$) and $\beta_{j}$ ($j=0,1,2,3$)
are given in Appendix \ref{sec:small angle expansion}.
Once the string configuration is fixed at each 
scattering event, we can immediately calculate 
the temperature deviation using Eq.(\ref{eq:small angle expansion of HSV}).
	
\subsubsection{Degrees of freedom of a string configuration}
	
Our remaining task is to determine string configurations.
A string configuration can be characterized by several random variables. 
Details about the random variables are discussed in 
Appendix~\ref{sec:random variables}. 
Here, we just count the number of degrees of freedom (dofs) to 
determine a string configuration.

For a string segment with the curvature taken into account,
originally we have 3 dof for ${\bm r}_0$, 6 dof for 
$\dot{\bm r}_0$ and ${\bm r}'_0$, 
and 9 dof for $\ddot{\bm r}_0$, $\dot{\bm r}'_0$ and ${\bm r}''_0$. 
For ${\bm r}_0$, using the coordinate dof, there remains only 1 dof,
which is the impact parameter $\delta$. 
For $\dot{\bm r}_0$ and ${\bm r}'_0$, 
the number of dof is reduced to 4 by using 2 constraint equations
(the gauge condition and the normalization) for the string dynamics.
The remaining dof are the magnitude of the velocity $v$ and 
the 3 angular parameters ($\theta$, $\phi$, $\psi$)
that determine the direction along the string and the velocity vector
in the plane orthogonal to the string direction. 
Finally, the number of dof of 
$\ddot{\bm r}_0$, $\dot{\bm r}'_0$ and ${\bm r}''_0$
is reduced to 4 using 3 equations of motion and 2 constraints.
One of them corresponds to
the inner product of the velocity vector and the curvature 
vector, $\dot{\bm r}_0\cdot{\bm r}''_0$.
The remaing 3 dof correspond to the curvature vector orthogonal
to $\dot{\bm r}_0$ and the rotation of the string segment
orthogonal to both $\dot{\bm r}_0$ and ${\bm r}''_0$.
Lacking the knowledge of how much freely moving string
segments would be curved, we set these remaing 3 dof to zero
for simplicity.
Thus we have 6 dof in total,
($\delta$, $v$, $\theta$, $\phi$, $\psi$, $\dot{\bm r}_0\cdot{\bm r}''_0$). 


\FIGURE[t]{\epsfig{file=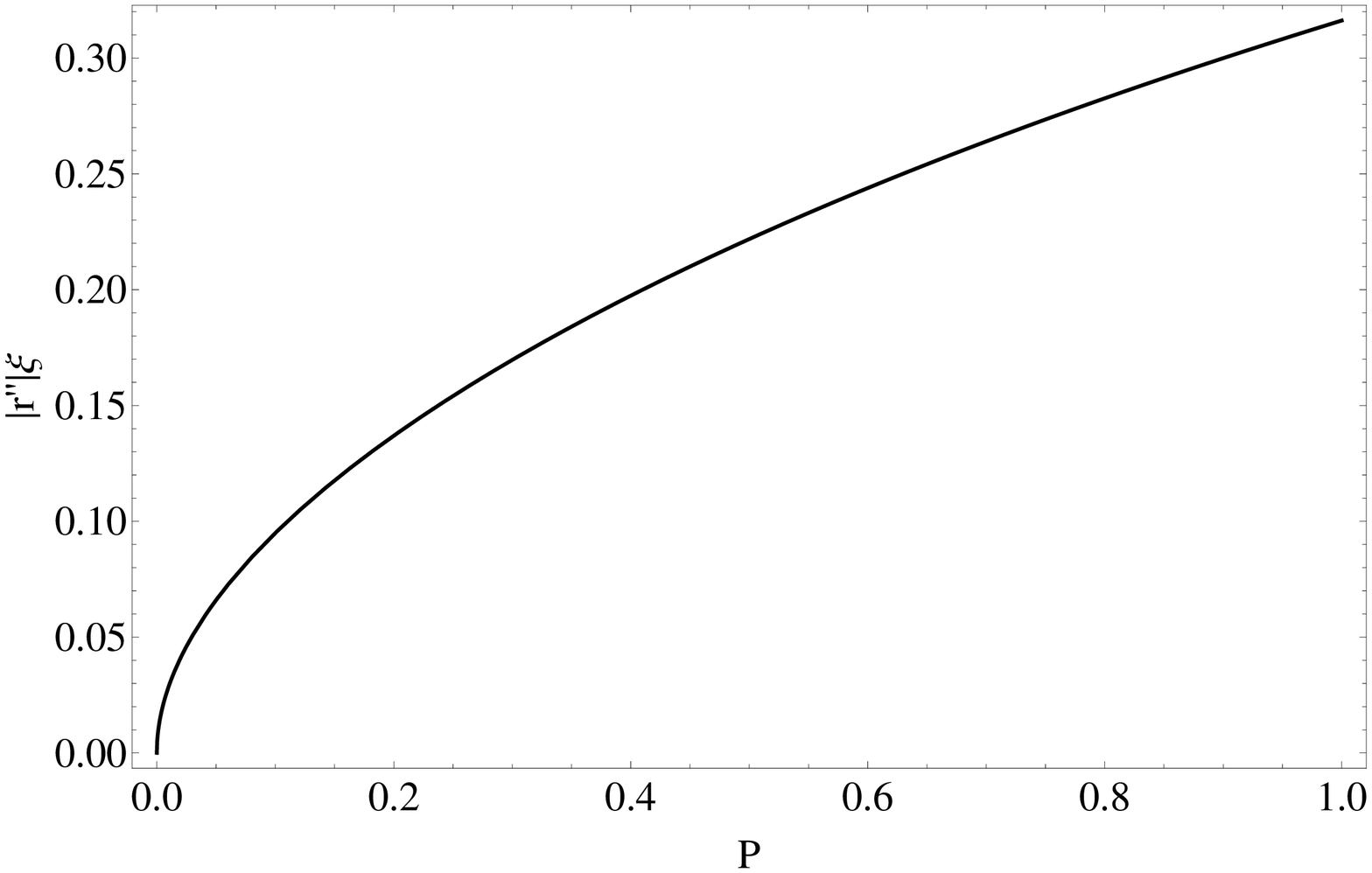,width=10cm} 
        \caption{The amplitude of the curvature vector normalized
by $\xi=1/H\gamma$, $|{\bm r}''_0|\xi =({\bm r}''_0\cdot\dot{\bm r}_0/v_{\rm rms})\xi
={\cal O}(\varepsilon)$ as a function of $P$.}%
	\label{fig:kMinus}}

In this paper, we further reduce the dof by
fixing the velocity and using the scaling assumption.
First we set $v=v_{\rm rms}$.
In the scaling regime, the curvature of a string segment 
and the velocity must be correlated.
Locally at each event of scattering, one may
neglect the Hubble expansion, and we have
\begin{eqnarray}
\frac{1}{a}\bigg\langle\dot{\bm r}\cdot\frac{1}{\epsilon}
\left(\frac{{\bm r}^{\prime}}{\epsilon}\right)^{\prime}\bigg\rangle
&\approx&
\left\langle\frac{\dot{\bm r}\cdot{\bm r}^{\prime\prime}}{a}
\right\rangle
\approx
\frac{\dot{\bm r}_0\cdot{\bm r}^{\prime\prime}_0}{a}
\approx 2Hv_{\rm rms}^2\left( 1-v_{\rm rms}^2\right)
\,.\label{eq:numerical curvature-velocity correlation}
\end{eqnarray}
Then
\begin{equation}
\dot{\bm r}_0\cdot{\bm r}^{\prime\prime}_0=
2aHv_{\rm rms}^2\left( 1-v_{\rm rms}^2\right)\,. 
\label{eq:reduced correlation}
\end{equation}
Thus, the remaining dof are ($\delta$, $\theta$, $\phi$, $\psi$).
For a demonstration of the temperature fluctuation due to a 
curved segment, we show the CMB map with 
$\theta =\phi =\pi /2$ and $\psi =0$ in Fig.~\ref{fig:HSV_map}.


\FIGURE[t]{\epsfig{file=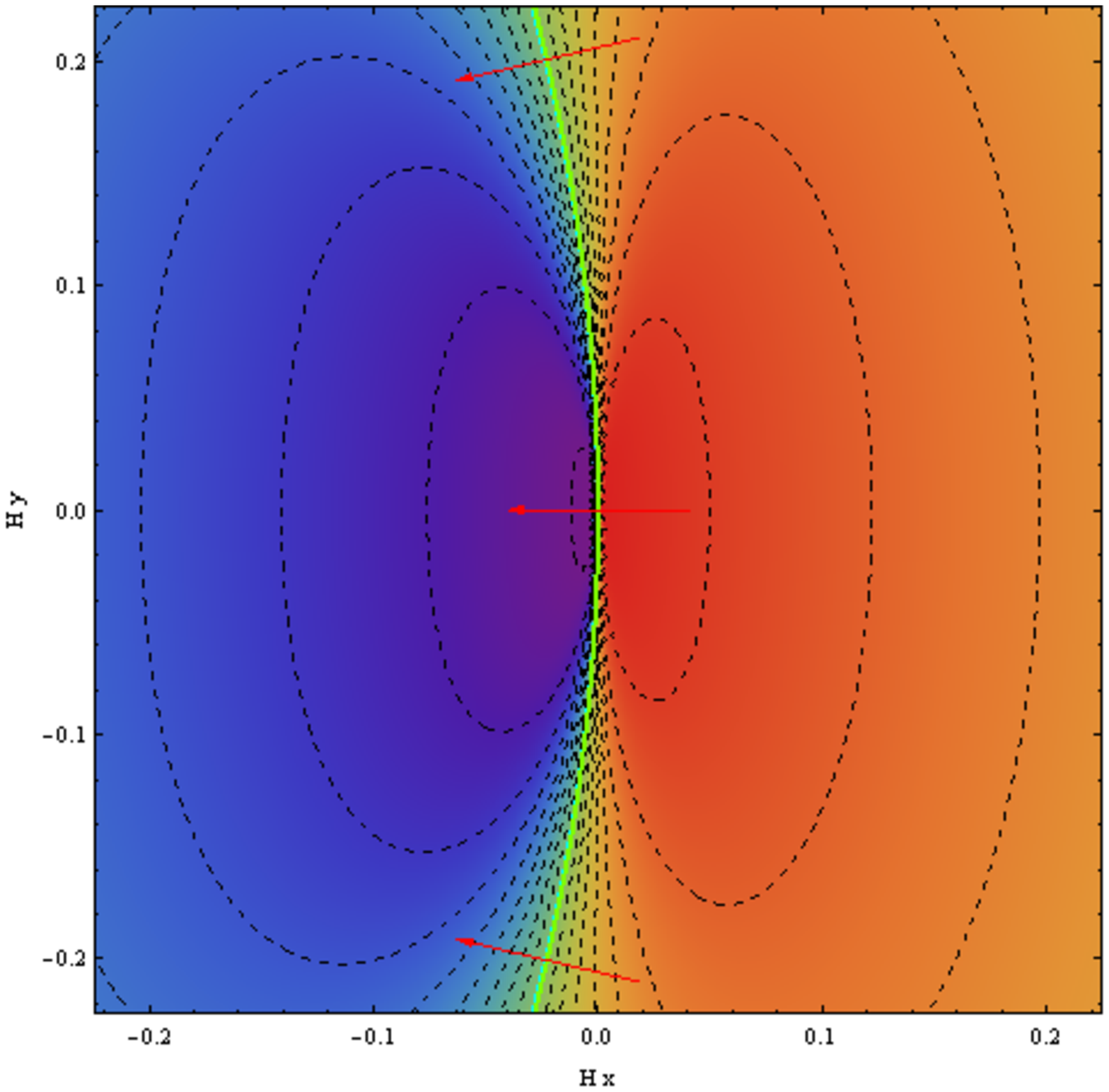,width=10cm} 
        \caption{The map of the temperature fluctuations due to a curved
string segment with $\theta =\phi =\pi /2$ and $\psi =0$.
Also we define $(Hx,Hy)$ as the Hubble normalized two-dimensional coordinate
relative to $\sigma =0$ in the small patch of sky.
The green line denotes the position of the string segment, 
${\bm X}^{\perp}(\sigma )$, 
the dashed lines are contour lines with the width $1G\mu$ and
the red vectors are the velocity vectors at each point.}%
	\label{fig:HSV_map}}

\subsection{Numerical Results}


\FIGURE[t]{\epsfig{file=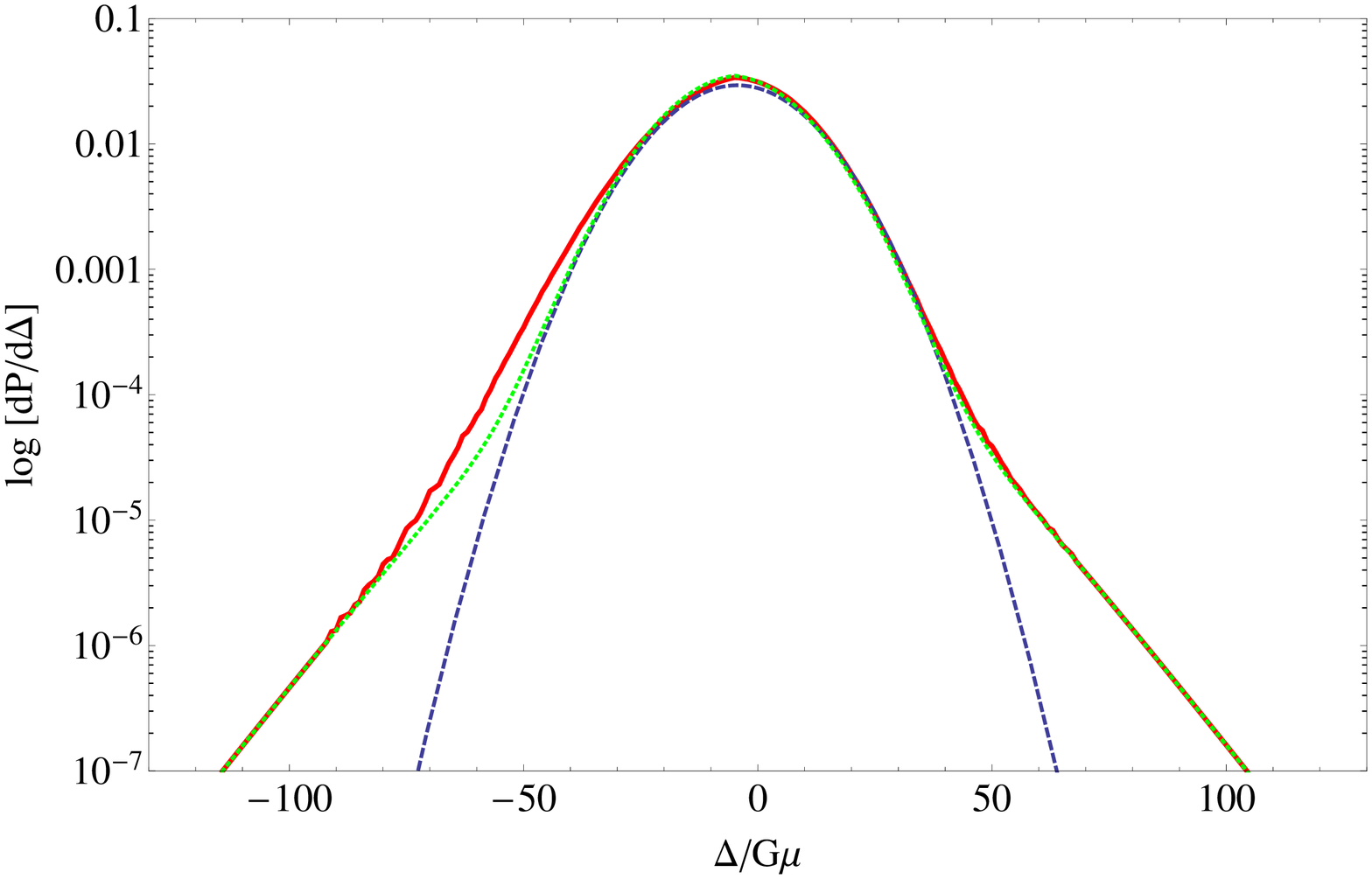,width=15cm} 
        \caption{The one-point pdf of the CMB temperature fluctuations 
induced by curved cosmic strings and kinks for
the intercommuting probability $P=1$. 
The solid curve (in red) is the one-point pdf obtained by averaging over
scattering of $10^7$ photons by curved string segments,
added on top of an analytic pdf for kinks obtained in \cite{Takahashi:2008ui}.
The dashed line (in blue) is the best Gaussian fit.
The dotted line (in green) represents the 
one-point pdf for straight segments and kinks.}%
	\label{fig:pdf_P1}}

We performed Monte Carlo simulations with 
$10^7$ photons on NEC Express 5800 XeonMP $3.16$GHz at 
Yukawa Institute for Theoretical Physics, Kyoto University.
It took two CPU days for $P=1$, three CPU days for $P=0.5$ 
and five CPU days for $P=0.1$. 

For $P=1$, the obtained one-point pdf 
of the total temperature fluctuations 
is shown in Fig.~\ref{fig:pdf_P1}.
In order to take into account the contribution from kinks, 
we just added an analytic pdf for kinks obtained in
our previous paper~\eqref{eq:deltakink}.
The one-point pdf significantly deviates from the best Gaussian fit.
The sample standard deviation is 
$\sigma_{\rm sim}(P=1) \approx 14 G\mu$, consistent with 
the analytical estimate in \cite{Takahashi:2008ui} and
Fig.~\ref{fig:sigma_seg}
as well as with the numerical result of \cite{Fraisse:2007nu}.
The sample skewness measured in Fig.~\ref{fig:pdf_P1} is 
\begin{eqnarray}
g_1(P=1)
=\frac{\overline{(\Delta -\bar{\Delta})^3}}{\sigma_{\Delta}^{3}}
\approx -0.14\,.
\end{eqnarray}
Thus the one-point pdf has a negative skewness.


\FIGURE[t]{\epsfig{file=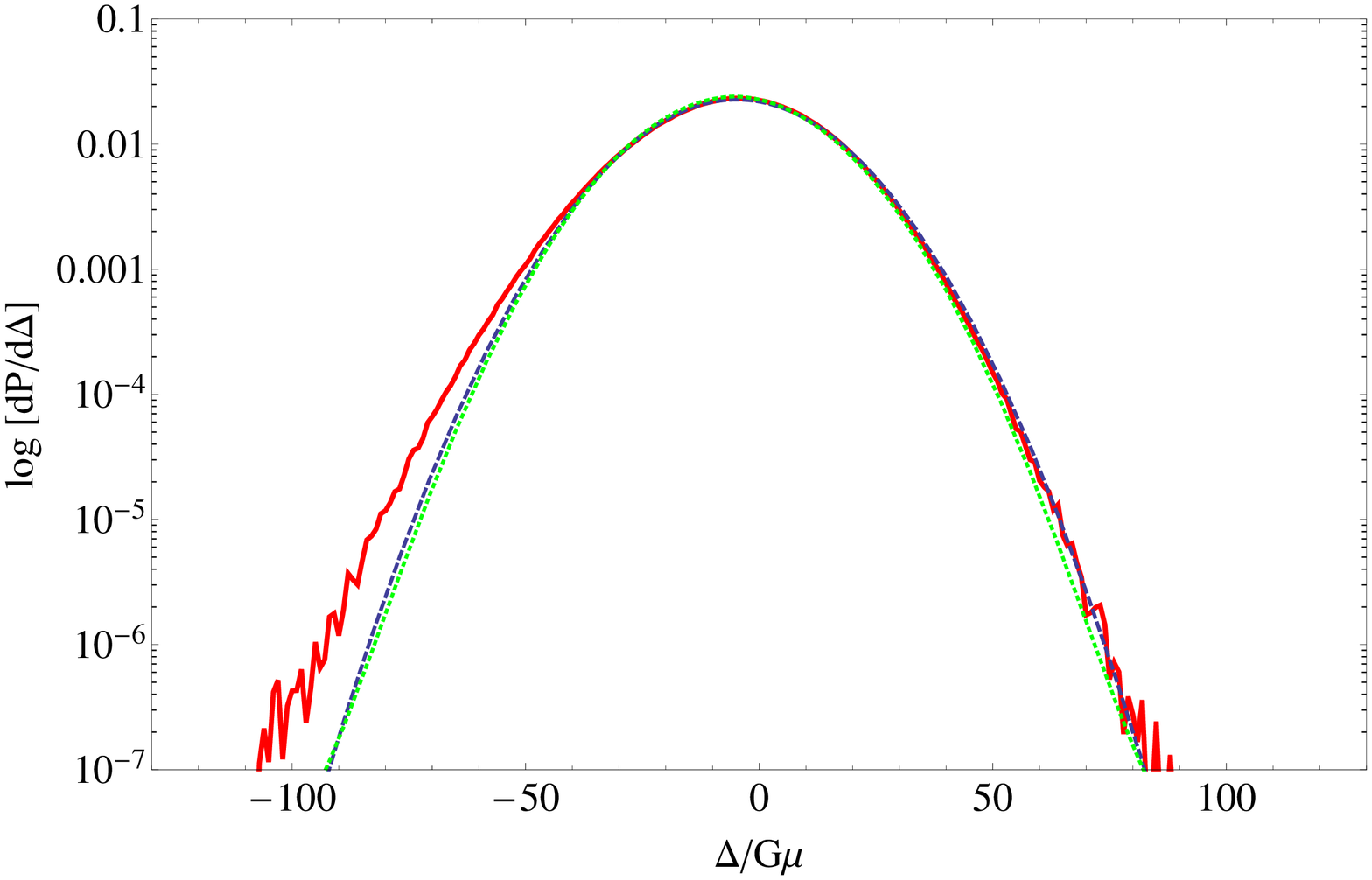,width=15cm} 
        \caption{The same as Fig.~4
but for the intercommuting probability $P=0.5$.}%
	\label{fig:pdf_P05}}


\FIGURE[t]{\epsfig{file=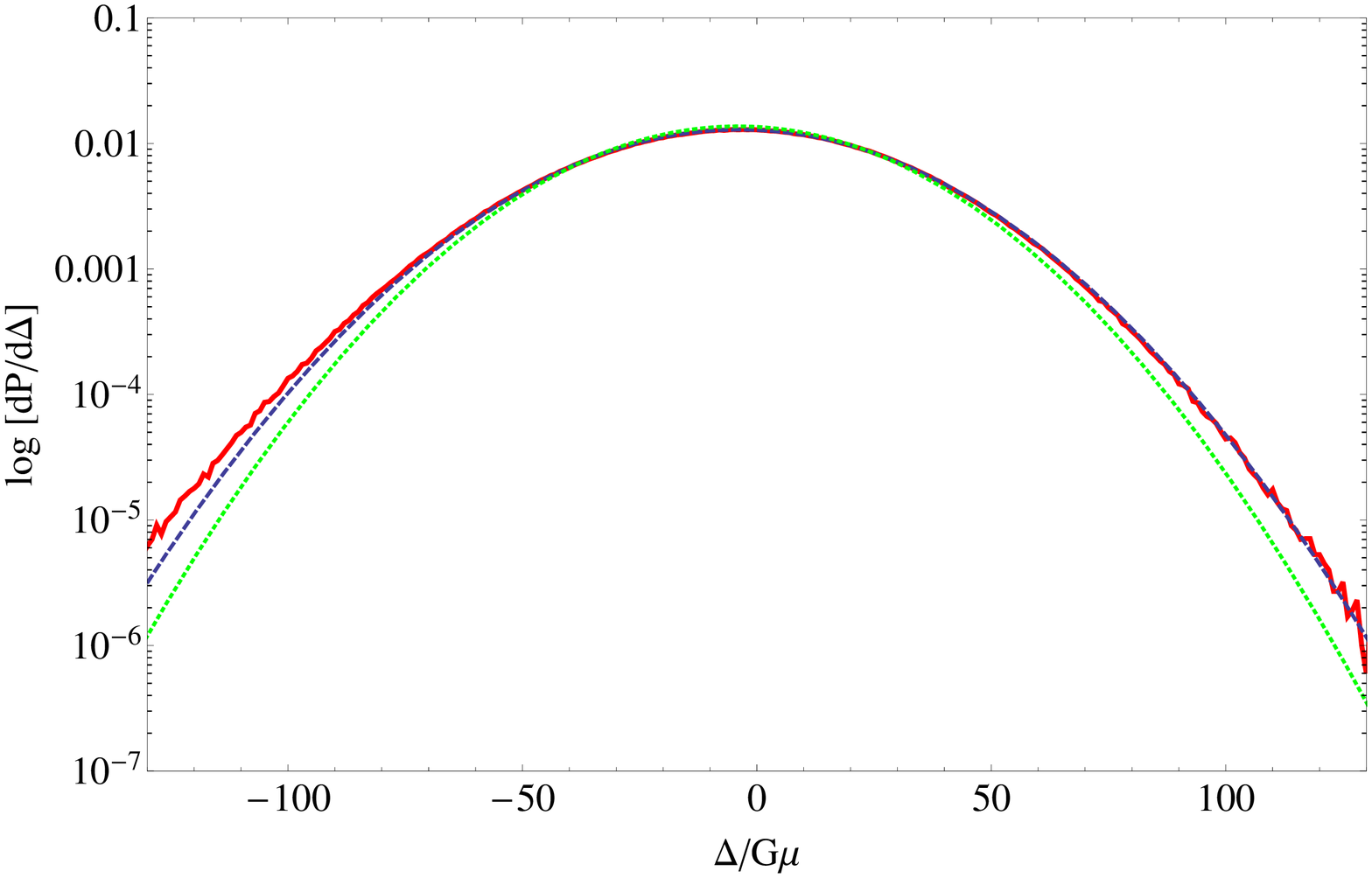,width=15cm} 
        \caption{The same as Fig.~4
but for the intercommuting probability $P=0.1$.}%
	\label{fig:pdf_P01}}

In order to investigate the dependence of the one-point pdf on
the intercommuting probability $P$, 
we computed the one-point pdfs for $P=0.5$ and $P=0.1$ and 
the obtained one-point pdfs are shown in 
Figs.~\ref{fig:pdf_P05} and \ref{fig:pdf_P01}.
The standard deviations are 
$\sigma_{\rm sim}(P=0.5)\approx 18G\mu$ for $P=0.5$ and
$\sigma_{\rm sim}(P=0.1)\approx 31G\mu$ for $P=0.1$. 
These are consistent with the analytical estimate
in \cite{Takahashi:2008ui} and Fig.~\ref{fig:sigma_seg}.
The sample skewness was found as
\begin{eqnarray}
g_1(P<1)=
\left\{
\begin{array}{ll}
-0.11 &\mbox{for}\ P=0.5\,,\\
-0.04 &\mbox{for}\ P=0.1\,.\
\end{array}
\right.
\end{eqnarray}
Furthermore, we calculated the skewness for various 
values of $P$ and obtained a power-series best fit 
in Fig.~\ref{fig:P-skewness}.
As we see, the dispersion of the Gaussian part
increases as $P$ decreases (see Fig.~\ref{fig:sigma_seg}) 
and the non-Gaussian features, i.e., the skewness
and the non-Gaussian tail, are suppressed.
Our result suggests that the sample skewness for a sufficiently
small $P$ approaches $g_1\lesssim \text{(a\ few)}\times 10^{-2}$.


\FIGURE{\epsfig{file=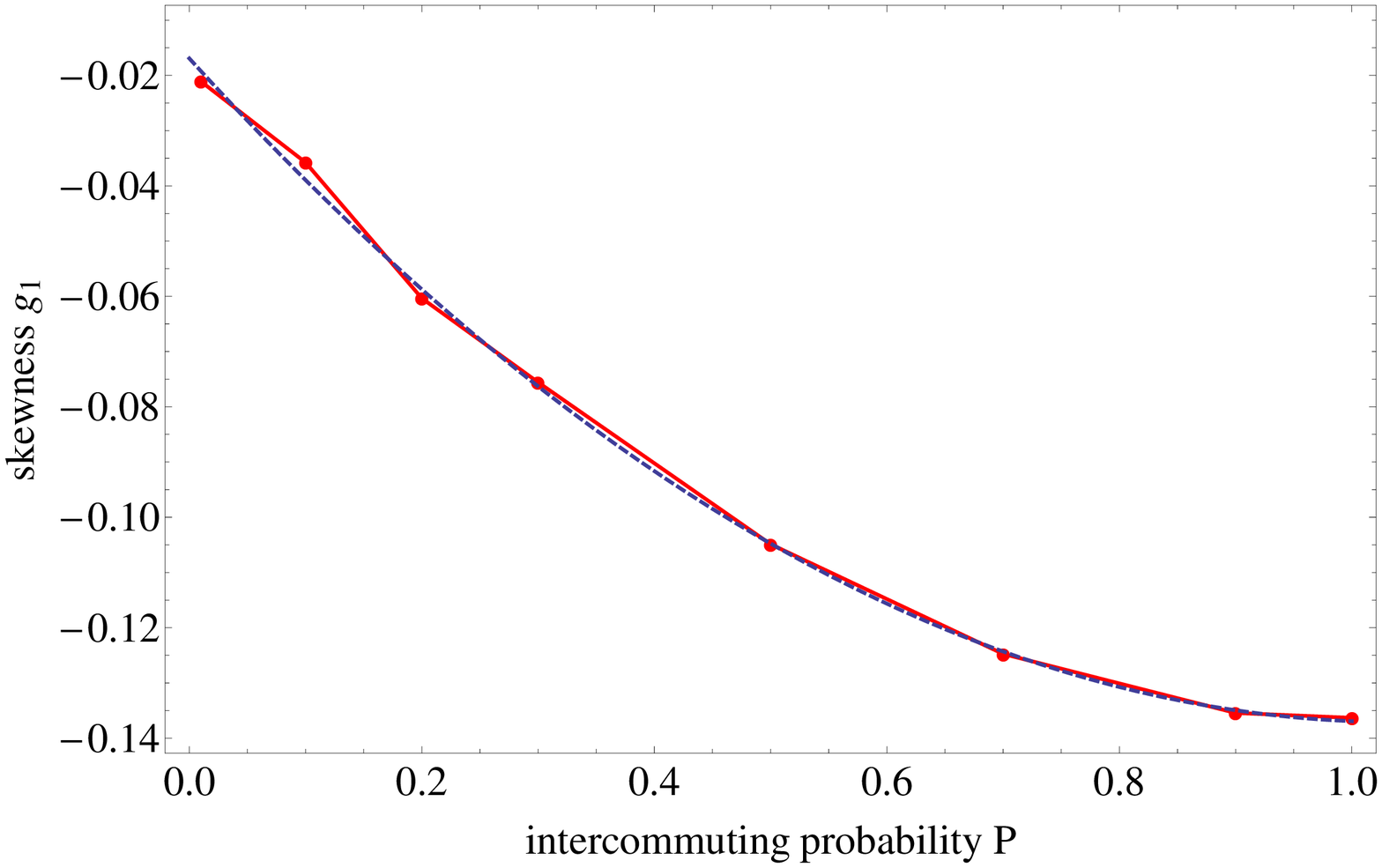,width=15cm} 
        \caption{The skewness as a function of the
intercommuiting probability $P$.
The solid dots (in red) are the skewness obtained
by simulations, each with $10^7$ photons.
The dashed line (in blue) is the power-series best fit 
function: $g_1(P)\approx 0.11P^2-0.23P-0.02$.}%
	\label{fig:P-skewness}}

\section{Summary}
\label{sec:summary}

In this paper, we have computed the one-point pdf of small-angle 
CMB temperature fluctuations due to the ``curved'' cosmic (super-)string
segments by performing Monte Carlo simulations.
Our purpose was to examine the numerical result obtained
in \cite{Fraisse:2007nu} with a simple model of the 
string network under a few reasonable assumptions,
i.e., the velocity-dependent one-scale (VOS) model. 
We have found that the presence of the nonzero momentum parameter $k$
corresponds to the correlation between the curvature and the velocity
of segments, and therefore the Hubble expansion is essential
for a nontrivial correlation (see Eq.~(\ref{eq:k def}) 
and Eq.~(\ref{eq:curvature-velocity correlation})).
We have also found another type of correlation 
between the curvature and the velocity induced by the effective 
curvature along the intersection of a string worldsheet and 
the observer's past light-cone (``light-cone effect'').
The light-cone effect always exists independent from
the properties of string configuration and 
the cosmic volume expansion.

Combining these two effects, we have calculated the one-point pdf
numerically and showed that it reproduces features of the numerical
simulations in \cite{Fraisse:2007nu} very well.
In particular, we have found that the one-point pdf has a
negative skewness. 
It was also found that as $P$ decreases
the standard deviation of the Gaussian part
increases and the non-Gaussian tails 
and the skewness are suppressed.
Our result suggests that, for the sufficiently small $P$, 
the skewness approaches $\lesssim \text{(a\ few)}\times 10^{-2}$.

Note that the observed temperature fluctuations
are given by contributions not only from
cosmic (super-)strings but also from 
primary and other secondary anisotropies.
It is known that conventional cosmic strings can contribute
at most $10\%$ of the total power at $\ell =10$
in the CMB fluctuations~\cite{Bevis:2007gh}.
Thus, in real world, contributions from primary 
anisotropies make the skewness significantly small
on large angular scales.
In order to suppress this effect of the large-scale anisotropies,
it would be effective to choose an observation area
of a sufficiently small angular size ($\sim O(10){\rm arcmin}^2$) 
where the primary CMB fluctuations have been completely damped out.
This would remove the contribution of the primary
fluctuations to the denominator of the skewness formula
(see Eq.~\eqref{eq:skew def}), hence enhance the observability
of the skewness.

It is known that ${\cal O}(0.1)$ skewness
would indicate nonlinear parameter 
$|f_{\rm NL}|\approx {\cal O}(10^3)$~\cite{Cayon:2002jq,Hindmarsh:2009qk}. 
In this paper, we have shown that, for smaller intercommuting 
probability $P$, non-Gaussian features become more difficult to find. 
To discuss the relation between the skewness and the non-Gaussian
parameter $f_{\rm NL}$ in the bispectrum, it is important to compute the 
angular power spectrum in our model. 
The work along this direction is in progress~\cite{Takahashi:future}.
Since the non-Gaussian features significantly depend on 
the intercommuting probability, they could help us to distinguish 
field theoretic cosmic strings and cosmic superstrings.

\section*{Acknowledgments}

DY thanks T. Azeyanagi, T.W.B. Kibble, K. Murata, A. Linde, Y. Sekino, 
T. Tanaka and V. Vanchurin for useful comments.
We thank the organizers and participants of ``The non-Gaussian Universe''
workshop at Yukawa Institute for Theoretical Physics for stimulating
discussion and presentations.
Numerical computation in this work was carried out on NEC $3.16$ GHz 
at Yukawa Institute for Theoretical Physics.
This work was supported in part by Monbukagaku-sho 
Grant-in-Aid for the Global COE programs
``The Next Generation of Physics, Spun from Universality 
and Emergence'' at Kyoto University and ``Quest for Fundamental Principles
in the Universe: from Particles to the Solar System and the Cosmos'' at
Nagoya University.
This work was also supported by JSPS Grant-in-Aid for Scientific Research 
(A) No.~18204024 and by Grant-in-Aid for Creative Scientific Research No.~19GS0219.
KT was supported by Grand-in-Aid for Scientific Research No.~21840028.
YS, DY and AN were supported by Grant-in-Aid for JSPS Fellows No.~19-7852, 
No.~20-1117 and No.~21-1899 respectively.

\appendix

\section{Reduction of the HSV formula}
\label{sec:small angle expansion}

The reduced HSV formula is
\begin{eqnarray}
\Delta =-4G\mu\int_{\Sigma}d\sigma
\frac{\Bigl[{\bm X}^{\perp}+\frac{(X^{\perp})^2}{2X}{\bm n}\Bigr]
\cdot\tilde{\bm u}}
{\Bigl( 1-\frac{\bm X}{X}\cdot\dot{\bm r}\Bigr) 
\left( X^{\perp}\right)^{2}}
\Biggl|_{\eta =\eta_{\rm lc}(\sigma )}
\,,
\end{eqnarray}
where we have introduced 
$\tilde{\bm u}=(1+{\bm n}\cdot\dot{\bm r}){\bm u}$,
and added terms second order in the small angle approximation.

We set
\begin{eqnarray}
{\bm X}^{\perp}(\sigma )
&=&\mbox{\boldmath $\delta $}
+\left(\frac{d{\bm X}^{\perp}}{d\sigma}\right)_{0}\sigma
+\frac{1}{2}\left(\frac{d^2{\bm X}^{\perp}}{d\sigma^2}\right)_{0}
\sigma^{2}+O(\varepsilon^3 X_0)\,,
\end{eqnarray}
where we have introduced the expansion parameter $\varepsilon$
such that $\delta/X_0=O(\varepsilon)$, etc., as discussed in
\ref{subsubsec:hsvformula}.
The absolute value of ${\bm X}^{\perp}$ is then expanded as
\begin{eqnarray}
&&\Bigl[ X^{\perp}(\sigma )\Bigr]^{2}=\delta^{2}
+2\Biggl[\mbox{\boldmath $\delta $}
\cdot\left(\frac{d{\bm X}^{\perp}}{d\sigma}\right)_{0}\Biggr]\sigma
+\Biggl[
\biggl|\frac{d{\bm X}^{\perp}}{d\sigma}\biggl|_{0}^2
+\mbox{\boldmath $\delta $}\cdot\left(\frac{d^2{\bm X}^{\perp}}
{d\sigma^2}\right)_{0}
\Biggr]\sigma^{2}
\nonumber\\
&&\ \ \ \ \ \ \ \ \ \ \ \ \ \ \ \ \ 
+\Biggl[\left(\frac{d{\bm X}^{\perp}}{d\sigma}\right)_{0}
\cdot\left(\frac{d^2{\bm X}^{\perp}}{d\sigma^2}\right)_{0}\Biggr]
\sigma^{3}+O(\varepsilon^4 X_0^2)
\nonumber\\
&&\ \ \ \ \ \ \ \ \ \ \ \ \ 
\equiv a_0+a_1\sigma +a_2\sigma^{2}+a_3\sigma^{3}+O(\varepsilon^4 X_0^2)
\,.
\end{eqnarray}


\FIGURE[t]{\epsfig{file=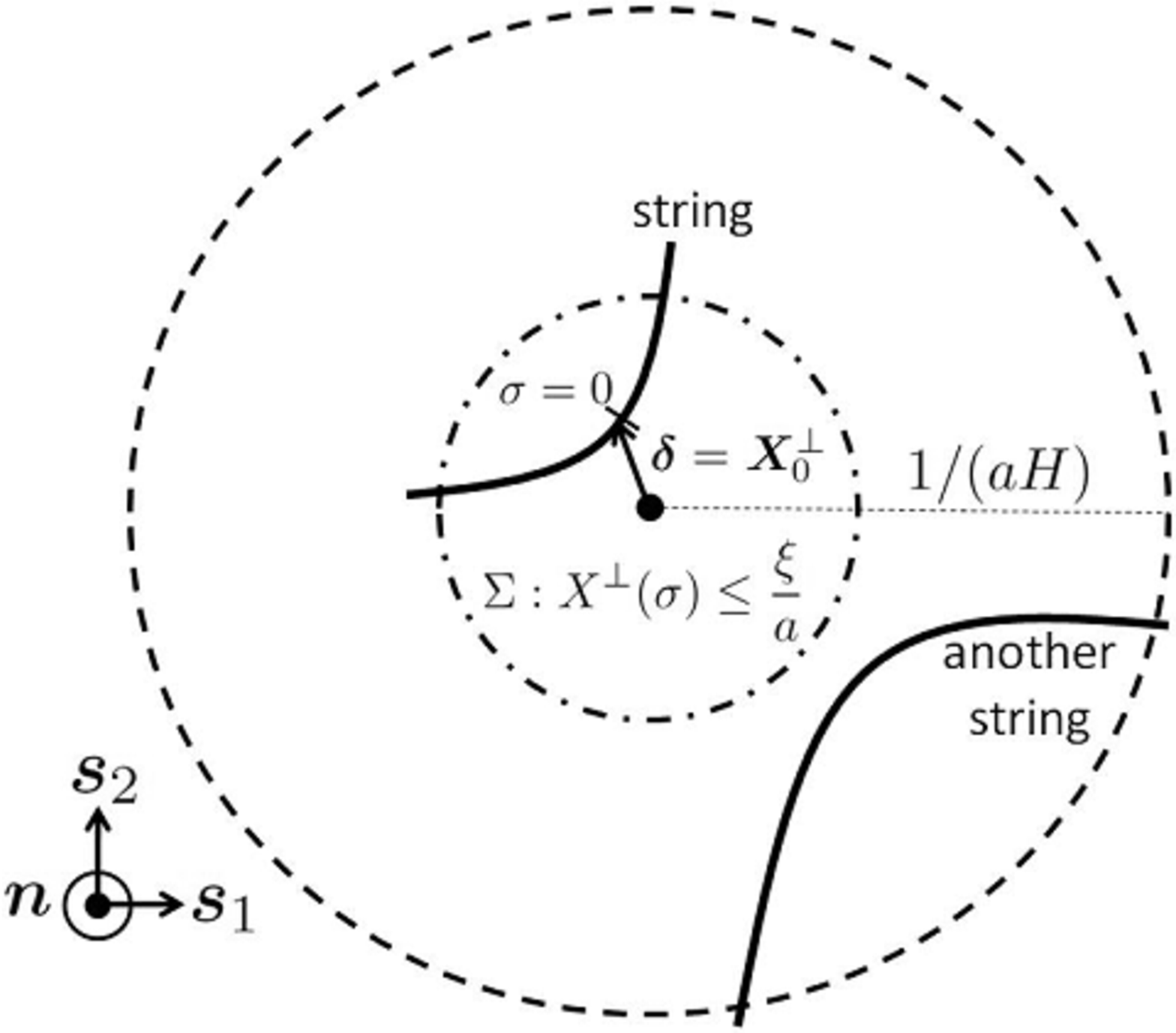,width=10cm} 
        \caption{A schematic picture of the surface perpendicular to 
the line of sight ${\bm n}$ at scattering.}%
	\label{fig:time_const_surface}}

We also have another useful expression, 
\begin{eqnarray}
&&1-\frac{\bm X}{X}\cdot\dot{\bm r}
\approx 1+{\bm n}\cdot\dot{\bm r}
-\frac{{\bm X}^{\perp}\cdot\dot{\bm r}}{X}
\equiv L_0+L_1\sigma +O(\varepsilon^2)\,,
\end{eqnarray}
where
\begin{eqnarray}
&&L_0=1+{\bm n}\cdot\dot{\bm r}_0
-\frac{\mbox{\boldmath $\delta $}\cdot\dot{\bm r}_0}{X_0}
\,,\\
&&L_1=
\left({\bm n}\cdot\frac{d\dot{\bm r}}{d\sigma}\right)_{0}
-\frac{1}{X_0}
\biggl[\left(\frac{d{\bm X}^{\perp}}{d\sigma}\right)_{0}\cdot\dot{\bm r}_0\biggr]
\,.
\end{eqnarray}
Then the HSV formula due to a curved segment
can be approximated as 
\begin{eqnarray}
&&\Delta= -4G\mu\int_{\Sigma} d\sigma
\frac{\sum_{i=0}^{2}\alpha_{i}\sigma^{i}}
{\sum_{j=0}^{3}\beta_{j}\sigma^{j}}
\,,
\end{eqnarray}
where the coefficients of the numerator are
\begin{eqnarray}
&&\alpha_{0}=\mbox{\boldmath $\delta $}\cdot\tilde{\bm u}_0
+\frac{\delta^{2}}{2X_0}({\bm n}\cdot{\bm u}_0)
\,,\\
&&\alpha_{1}=\mbox{\boldmath $\delta $}
\cdot\left(\frac{d\tilde{\bm u}}{d\sigma}\right)_{0}
+\left(\frac{d{\bm X}^{\perp}}{d\sigma}\right)_{0}\cdot\tilde{\bm u}_0
+\frac{1}{2X_0}\Biggl\{
a_1({\bm n}\cdot\tilde{\bm u}_0)
+a_0\biggl[{\bm n}\cdot\left(\frac{d\tilde{\bm u}}{d\sigma}\right)_{0}
\biggr]\Biggr\}
\,,\\
&&\alpha_{2}=\left(\frac{d{\bm X}^{\perp}}{d\sigma}\right)_{0}
\cdot\left(\frac{d\tilde{\bm u}}{d\sigma}\right)_{0}
+\frac{1}{2}\biggl[
\left(\frac{d^2{\bm X}^{\perp}}{d\sigma^2}\right)_{0}
\cdot\tilde{\bm u}_0
\biggr]
\,,
\end{eqnarray}
and those of the denominator are
\begin{eqnarray}
&&\beta_{0}=L_0a_0
\,,\\
&&\beta_{1}=L_0a_1+L_1a_0
\,,\\
&&\beta_{2}=L_0a_2+L_1a_1
\,,\\
&&\beta_{3}=L_0a_3+L_1a_2\,.
\end{eqnarray}
A schematic picture of the relation between the normalized bases
and the origin of the worldsheet coordinates
is given in Fig.~\ref{fig:time_const_surface}.

\section{Random Variables}
\label{sec:random variables}

For our Monte Carlo simulations, we need to specify 
random variables in a physically reasonable way. 
In our situation, the random variables are 
those which specify the string configuration at each scattering event. 
They are the string position ${\bm r}_0$, 
the velocity $\dot {\bm r}_0$, the direction ${\bm r}_0'$, 
the curvature ${\bm r}_0''$, 
the rotation $\dot{\bm r}_0'$ and the acceleration ${\bm r}_0''$.
Thus there are 18 parameters in total. 
However, not all of them are independent from each other
because of the equations of motion and constraint equations. 

Since at each scattering event, the cosmic expansion can be
neglected, the equations of motion can be approximated by those on the 
Minkowski background.
In the conformal and temporal gauge, 
the equations of motion are given by 
\begin{equation}
\ddot{\bm r}-{\bm r}''=0. 
\label{eq:eominm}
\end{equation}
The gauge conditions lead constraint equations as
\begin{eqnarray}
&&\dot{\bm r}\cdot {\bm r}'=0 
\,,\
\dot{\bm r}^2+{{\bm r}'}^2=1. 
\label{eq:constinm1}
\end{eqnarray}
Differentiating the constraint equations, we have
\begin{eqnarray}
&&{\bm r}_0^{\prime\prime}\cdot\dot{\bm r}_0
=-{\bm r}^{\prime}_0\cdot\dot{\bm r}^{\prime}_0
\,,\ 
\ddot{\bm r}_0\cdot{\bm r}^{\prime}_0
=-\dot{\bm r}_0\cdot\dot{\bm r}^{\prime}_0
\,.
\label{eq:constinm2}
\end{eqnarray}

Using Eqs.~\eqref{eq:eominm}-\eqref{eq:constinm2}, 
we can derive the following expressions:
\begin{eqnarray}
&&{\bm r}^{\prime}_0 
=\sqrt{1-v^2}{\bm n}_{\parallel}
\,,\ 
\dot{\bm r}_0
=v{\bm t}_{1}
\,,\\
&&\ddot{\bm r}_0={\bm r}^{\prime\prime}_0
\approx
-\frac{v}{\sqrt{1-v^2}}
(\dot{\bm r}^{\prime}_0\cdot{\bm t}_{1}){\bm n}_{\parallel}
+({\bm r}^{\prime\prime}_0\cdot{\bm t}_{1}){\bm t}_{1}
+({\bm r}^{\prime\prime}_0\cdot {\bm t}_{2})
{\bm t}_{2}
\,,\\
&&\dot{\bm r}^{\prime}_0
\approx 
-\frac{v}{\sqrt{1-v^2}}
({\bm r}^{\prime\prime}_0\cdot{\bm t}_{1}){\bm n}_{\parallel}
+(\dot{\bm r}^{\prime}_0\cdot{\bm t}_{1}){\bm t}_{1}
+(\dot{\bm r}^{\prime}_0\cdot{\bm t}_{2})
{\bm t}_{2}\,,
\end{eqnarray}
where $({\bm n}_{\parallel},{\bm t}_{1},{\bm t}_{2})$ are
unit basis vectors with ${\bm n}_{\parallel}$ parallel to ${\bm r}'_0$
and ${\bm t}_{1}$ parallel to $\dot{\bm r}_0$.
In terms of coordinate basis vectors $({\bm n},{\bm s}_{1},{\bm s}_{2})$
with ${\bm n}$ being the unit vector along the light of sight,
those parameters can be written as
\begin{eqnarray}
&&v=|\dot {\bm r}_0|
\,,\\
&&{\bm n}_{\parallel}=\sin\theta\cos\phi{\bm s}_{1}
+\sin\theta\sin\phi{\bm s}_{2}
+\cos\theta{\bm n}
\,,\\
&&{\bm t}_{1}
=(-\sin\psi\cos\theta\cos\phi
-\cos\psi\sin\phi ){\bm s}_{1}
\nonumber\\
&&\ \ \ \ \ \ 
+(-\sin\psi\cos\theta\sin\phi
+\cos\psi\cos\phi ){\bm s}_{2}
+\sin\psi\sin\theta{\bm n}
\,,\\
&&{\bm t}_{2}
=(-\cos\psi\cos\theta\cos\phi
+\sin\psi\sin\phi ){\bm s}_{1}
\nonumber\\
&&\ \ \ \ \ \ 
+(-\cos\psi\cos\theta\sin\phi
-\sin\psi\cos\phi ){\bm s}_{2}
+\cos\psi\sin\theta{\bm n}
\,,
\end{eqnarray}
where the ranges of the angular coordinates are
\begin{eqnarray}
0\leq \theta\leq \pi\,,
\quad
0\leq\phi\leq 2\pi\,,
\quad
0\leq\psi\leq 2\pi\,.
\end{eqnarray}
Thus the total number of degrees of freedom (dof) is reduced 
to 9; $\bigl(\delta$, $v$, $(\dot{\bm r}^{\prime}_0\cdot{\bm t}_{1})$, 
$({\bm r}''_0\cdot{\bm t}_{1})$, $(\dot{\bm r}'_0\cdot{\bm t}_{2})$, 
$({\bm r}''_0\cdot{\bm t}_{2})$, $\theta$, $\phi$, $\psi\bigr)$. 

The impact parameter $\mbox{\boldmath $\delta $} ={\bm X}^{\perp}_0$ has 
to be chosen randomly at each scattering.
The impact parameter has two dof.
One is the absolute value $\delta$ and another is the angular
coordinate $\theta_{X}$. Then
\begin{eqnarray}
&&{\bm X}^{\perp}_0=\delta\Bigl[
\cos\theta_{X}{\bm s}_{1}
+\sin\theta_{X}{\bm s}_{2}
\Bigr]\,.
\end{eqnarray}
Performing a 2-dimensional coordinate transformation,
\begin{eqnarray}
\left(
\begin{array}{c}
{\bm s}_{1}\\
{\bm s}_{2}\\
\end{array} 
\right)
=\left(
\begin{array}{cc}
\cos\chi & \sin\chi \\
-\sin\chi & \cos\chi \\
\end{array} 
\right)
\left(
\begin{array}{c}
\tilde{\bm s}_{1}\\
\tilde{\bm s}_{2}\\
\end{array} 
\right)\,,
\end{eqnarray}
the impact parameter ${\bm X}^{\perp}_0$ can by expressed as
\begin{eqnarray}
&&{\bm X}^{\perp}_0=\delta\Bigl[
\cos (\theta_{X}+\chi )\tilde{\bm s}_{1}
+\sin (\theta_{X}+\chi )\tilde{\bm s}_{2}
\Bigr]\,.
\end{eqnarray}
Therefore, by using the dof for the rotation of the 2-dimensional 
coordinate basis the angular dof $\theta_{X}$ can be chosen freely.
We set $\theta_{X}=\pi /2$.

 For the distribution of $\delta$, from its definition
we set $P(\delta)d\delta\propto\delta d\delta$ with the
cutoff at $\delta=\xi/a$. 
Then we fix the range of integration $X^{\perp}(\sigma )\leq \xi /a$
for a given $\delta$.
For a string segment with the parabola-type shape,
Eq.~(\ref{eq:parabola shape}), the condition can be written as 
\begin{eqnarray}
a_0-\frac{\xi^{2}}{a^2}+a_1\sigma +a_2\sigma^{2}+a_3\sigma^{3}
\leq 0\,.
\end{eqnarray}
For a given value of the impact parameter at $\delta \leq \xi/a$,
the range of $\sigma$ is chosen to satisfy the above
equation.

For the angular variables $\theta$, $\phi$ and $\psi$,
we choose uniform distributions 
$P(\cos\theta )=1/2\,, P(\phi)=P(\psi)=1/2\pi$.
As for $v$, we fix it at $v=v_{\rm rms}$ for simplicity. 

We fix the value of $({\bm r}^{\prime\prime}_0\cdot{\bm t}_{1})$ 
by using Eq.~(\ref{eq:numerical curvature-velocity correlation})
so that it is consistent with the correlation between the 
string velocity and the curvature. 
As for the remaining three dof, we set
$({\bm r}^{\prime\prime}_0\cdot{\bm t}_{2})=0$,
$(\dot{\bm r}^{\prime}_0\cdot{\bm t}_{1})=0$ and
$(\dot{\bm r}^{\prime}_0\cdot{\bm t}_{2})=0$. 
Note that this is solely due to the lack of our knowledge
about the curvature and rotation of string segments.


\end{document}